\documentclass{mem}
\usepackage{natbib}\usepackage{txfonts}\usepackage{balance}
\usepackage{graphicx}
\usepackage[a4paper,breaklinks,dvipdfm]{hyperref}
\idline{75}{282}
\begin{document}
\def\teff{$T\rm_{eff }$}
\def\kms{$\mathrm {km s}^{-1}$}

\title{
The multifrequency behaviour of the \\ recurrent nova RS Ophiuchi
}

   \subtitle{}

\author{
Val\'erio A. R. M. Ribeiro\inst{1}
          }

  \offprints{V. A. R. M. Ribeiro}

\institute{
Astrophysics, Cosmology and Gravity Centre,
Department of Astronomy,
University of Cape Town,
Private Bag X3,
Rondebosch 7701, South Africa 
\email{vribeiro@ast.uct.ac.za}
}

\authorrunning{Ribeiro}

\titlerunning{RS Ophiuchi}

\abstract{
This review concentrates on the multifrequency behaviour of RS Ophiuchi and in particular during its latest outburst. Confirmation of the 1945 outburst, bipolar outflows and its possible fate as a Type Ia Supernova are discussed.
}
\maketitle{}

\section{Introduction}
\begin{figure}[]
\resizebox{\hsize}{!}{\includegraphics[clip=true]{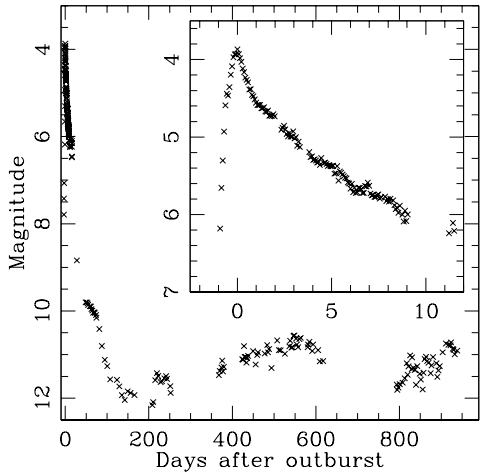}}
\caption{\footnotesize
Liverpool Telescope and SMEI light curves of the 2006 outburst \citep[from][]{DHB08,HBH10}. {\it Inset:} Detailed SMEI light curve of the outburst. See text for details.
}
\label{fig:outburst}
\end{figure}

\footnotetext[1]{SA SKA Fellow}RS Ophiuchi is a symbiotic recurrent nova (RN) with outbursts recorded in 1898, 1933, 1958, 1967, 1985 \citep{R87,RI87}, and its latest on 2006 February 12.83--12.94 \citep[Figure~\ref{fig:outburst};][]{NHK06,EBO08,HBH10}. However, during the period 1898--1933, before the star was known to be a RN, some outbursts might have occurred, most likely in 1907 \citep{S04} and the suggested 1945 outburst \citep{OM93} was confirmed by \citet[][see later]{AES11}.

The system, at a distance of 1.6$\pm$0.3 kph \citep{B87,BMS08}, comprises a massive white dwarf (WD) primary, probably close to the Chandrasekhar limit \citep[e.g.,][]{BQM09}, and red-giant secondary of spectral type estimated around M0/2III \citep[e.g.,][]{AM99}. Accretion of hydrogen-rich material from the red-giant onto the WD surface leads to a thermonuclear runaway (TNR), as in the outbursts of classical novae (CNe). The much shorter inter-outburst period for RNe compared to CNe is predicted to be due to a combination of high WD mass and a high accretion rate \citep[e.g.,][]{SST85,YPS05}. This paper reviews some of the results that have come from the study of RS Oph at a variety of wavelengths and in particular during its latest outburst.

\section{Multifrequency behaviour}

\subsection{Radio}
Radio observations in 2006 around day 13.8 after outburst with Very Long Baseline Interferometry (VLBI) showed a partial ring of non-thermal emission (from the expanding shock) which later developed in to a bipolar structure \citep{OBP06,OBB08}. The asymmetry was suggested to be due to absorption in the overlying red-giant wind and they noted more extended components emerging to the east and west.

The remnant was followed between days 34 -- 51 after outburst with the Very Long Baseline Array (VLBA) by \citet{SRM08} who found a central thermally dominated source linked by what appeared to be collimated non-thermal outflow which were interpreted as the working surfaces of jets. Jet collimation could for example be due to the expected accretion disc around the WD. It should be noted that Taylor et al. (1989) had also interpreted their 5 GHz VLBI map from day 77 after the 1985 outburst in terms of a central thermal source with expanding non-thermal lobes.

\subsection{Optical} 
\begin{figure}[]
\resizebox{\hsize}{!}{\includegraphics[clip=true]{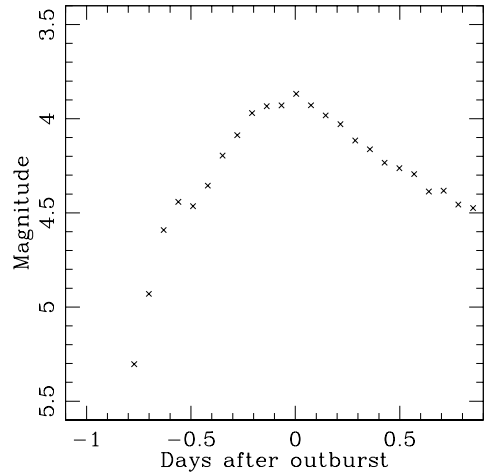}}
\caption{
\footnotesize
SMEI light curve detailing the initial outburst up to maximum \citep[from][]{HBH10}.
}
\label{fig:smei}
\end{figure}

The optical regime is perhaps one that has the greatest wealth of information published, mainly due to the accessibility to telescopes/instruments. It is at this wavelength that the nova in outburst is first detected, leading to triggers in other wavelengths. The outbursts of novae are generally caught near maximum or while it is on the decline. By combining all-sky surveys, similar to the Solar Mass Ejection Imager (SMEI), we can sample well the pre-maximum and early decline light curves of the outburst and here find interesting features (see Figures~\ref{fig:outburst} and \ref{fig:smei}). For example, \citet{HBH10} were able to observe the pre-maximum halt, at around day $-$0.5, which lasted for a few hours. They go further to suggest that the halt may be a temporary reversal in the light curve and may be related to a change in the mass loss rate, although its cause is poorly determined.

\begin{figure}[]
\resizebox{\hsize}{!}{\includegraphics[clip=true]{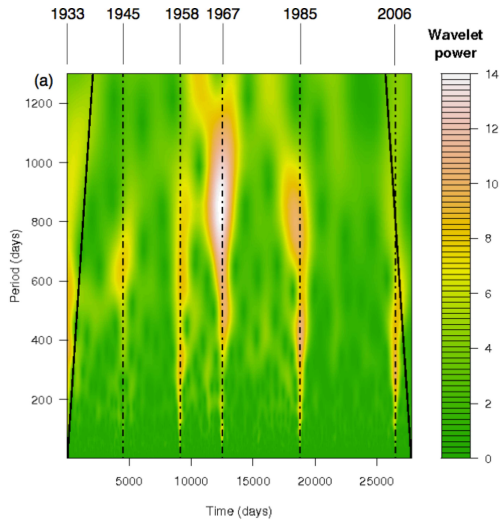}}
\caption{
\footnotesize
Wavelet power plot of the 1933 $-$ 2007 light curves, clearing showing all outburst dates (marked on top) and a pre-outburst signal which remains even in the absence of the outburst \citep[from][]{AES11}.
}
\label{fig:wave}
\end{figure}

\citet{AES11} examined the optical light curve between 1933 and 2007 obtained from the AAVSO database, which covered 6 outbursts (previously to publishing their results only 5 outburst were confirmed and one was suspected). By means of wavelet analysis they were able to confirm the suspected 1945 outburst and went further to suggest the existence of a pre-outburst signal starting up to a couple hundred days before the TNR (Figure~\ref{fig:wave}).

\begin{figure*}[]
\resizebox{\hsize}{!}{\includegraphics[clip=true]{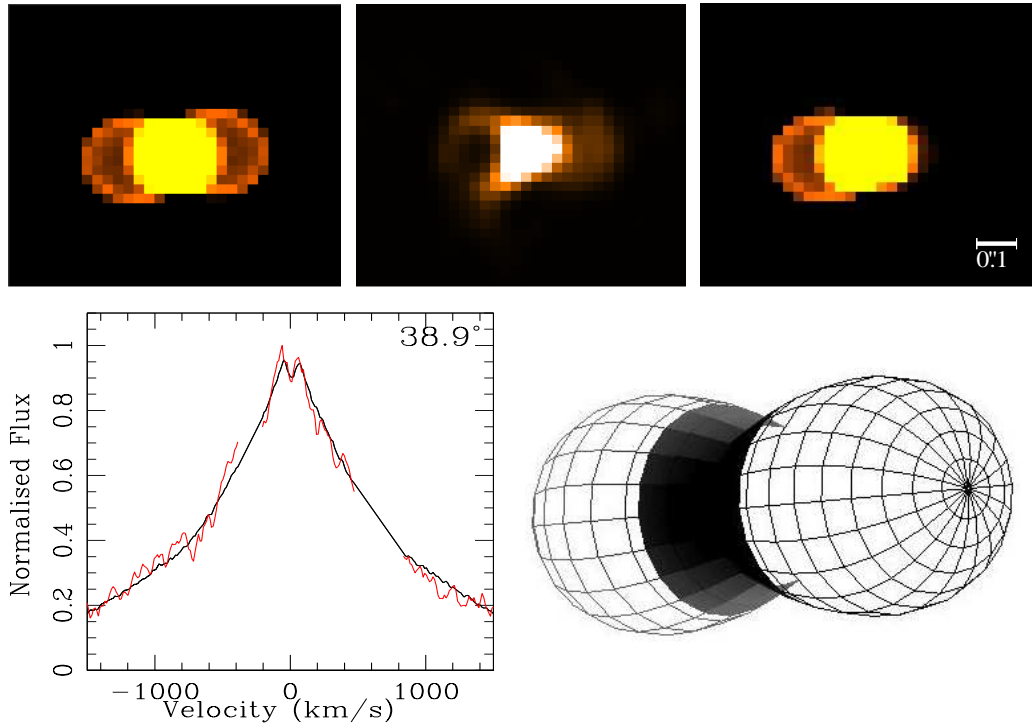}}
\caption{
\footnotesize
Top: synthetic image without the {\it HST} F502N ACS/HRC filter profile applied (left), enlarged ACS/HRC image at $t$ = 155 days after outburst (middle) and synthetic image with the ACS/HRC F502N filter profile applied (right). Bottom: best fit synthetic spectrum (black) overlaid with the observed spectrum (red). To the right is the model structure for RS Oph (outer dumbbell and inner hour glass). Images reproduced from \citet{RBD09,RBD11}.
}
\label{fig:hst}
\end{figure*}

{\it Hubble Space Telescope} ({\it HST}) imaging and ground-based spectroscopic observations at 155 days after outburst allowed \citet{RBD09} to model the ejecta as bipolar, at least in the ACS/HRC F502N {\it HST} filter, composed of an outer dumbbell and inner hour glass structures (Figure~\ref{fig:hst}). The inner hour glass was required as an over-density in order to replicate the observed [O~{\sc iii}] 5007\AA\ line profile. We were able to show that the observed apparent asymmetry of the ejecta in the {\it HST} image was an observational effect due to the finite width and offset of the central wavelength from the line centre of the F502N filter. From detailed kinematical modelling, the inclination of the system was determined as 39$^{+1}_{-9}$ degrees and maximum expansion velocity 5100$^{+1500}_{-100}$ km s$^{-1}$ (the range in velocity arises from the 1$\sigma$ errors on the inclination -- see  Figure~\ref{fig:hst}). This asymmetry was proposed to be due to interaction of the ejecta with a pre-existing red-giant wind \citep{BHO07,RBD09}.

The model was then evolved to 449 days after outburst, to match the second {\it HST} observations. However, in this case, due to the lack of simultaneous ground-based spectroscopy the model was harder to constrain. \citet{RBD09} suggested that at this time the outer dumbbell structure expanded linearly while the inner hour glass structure showed some evidence for deceleration. However, the results are open to over-interpretation due to the poor quality of the images at this epoch.

\citet{PCP11} using multi-epoch, high-resolution spectroscopy before, during and after the outburst studied the time evolution of Ca~{\sc ii}, Na~{\sc i} and K~{\sc i} absorption features and related these to density enhancements in the circumstellar material. They proposed a link between RS Oph and the Type Ia SN 2006X as a result \citep{PCC07}.

\subsection{X-ray}
\begin{figure}[]
\resizebox{\hsize}{!}{\includegraphics[clip=true]{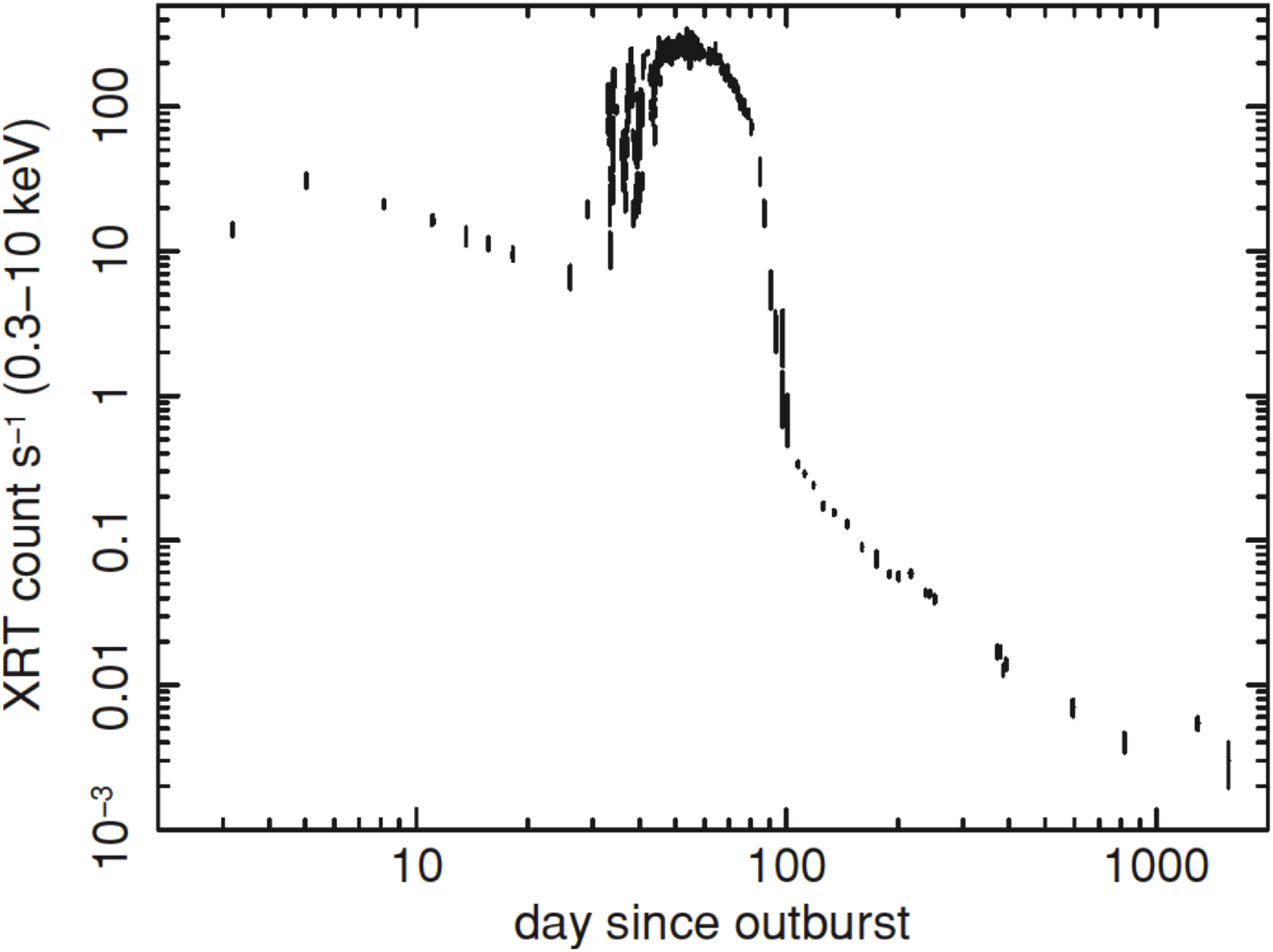}}
\resizebox{\hsize}{!}{\includegraphics[clip=true]{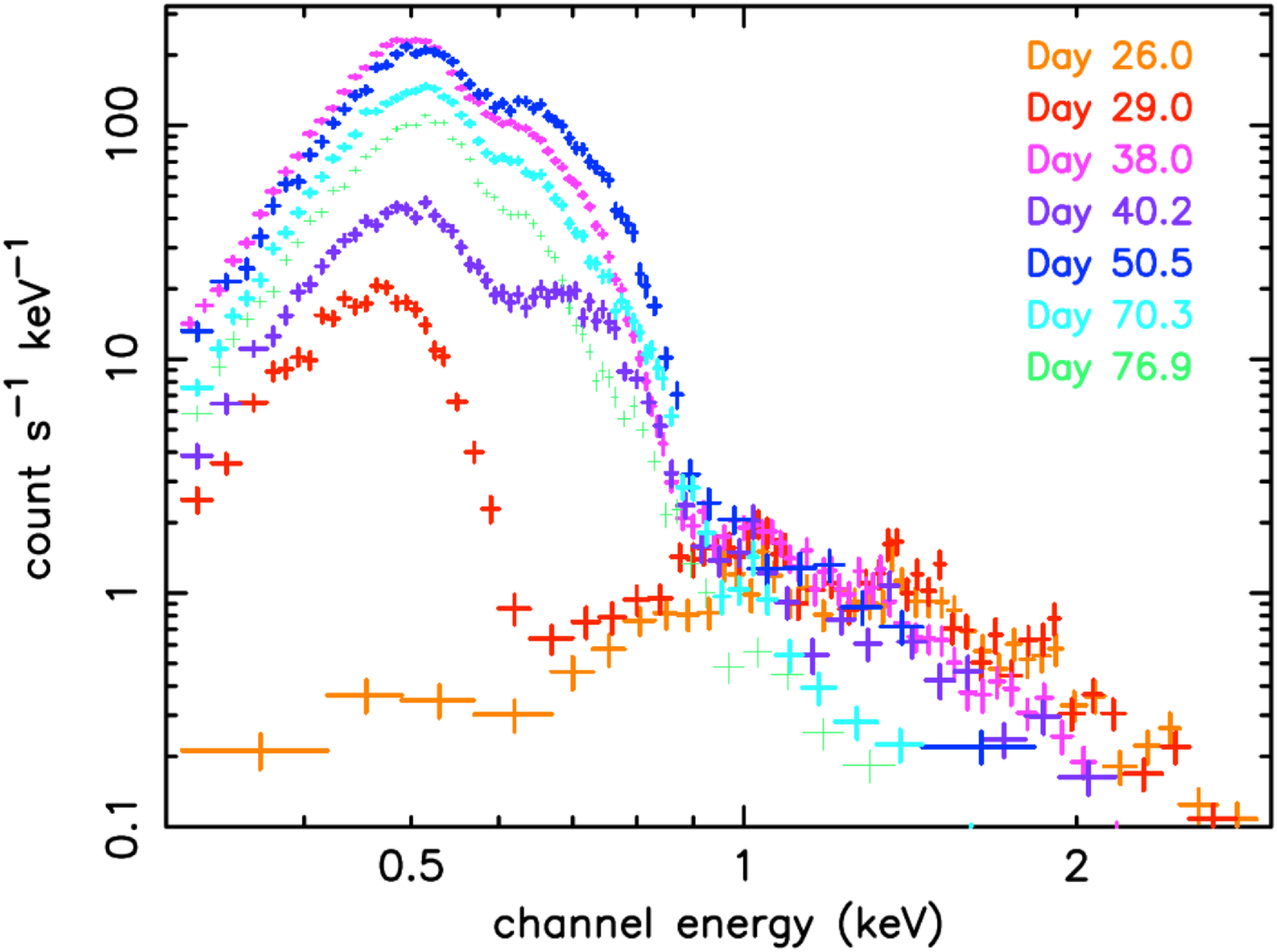}}
\caption{
\footnotesize
Top: {\it Swift}-XRT light curve of RS Oph. The super-soft source emerges around day 29, turning off around day 90 after outburst. Bottom: X-ray spectra detailing the evolution of the super-soft source. Images from  \citet{OPB11}.
}
\label{fig:swift}
\end{figure}

The initial decline, observed in Figure~\ref{fig:swift} (top), of the {\it Swift} X-ray count rates up to day 26 is well described by the evolution of shock systems as the outburst ejecta run into the pre-existing red-giant wind \citep{BOO06,SLM06}.The super-soft source then emerges by day 29 (Figure~\ref{fig:swift}, bottom), which samples nuclear burning on the WD surface, evident until about day 90 \citep{OPB11}.

The outburst was also followed in detail with X-ray grating observations with {\it Chandra} and {\it XXM-Newton} where two distinct temperature components were found which decayed with time \citep{NDS09}. The reduction in flux of these two components, hot and cooler, are consistent with radiative cooling and the expansion of the ejecta, respectively \citep{NDS09}. {\it Chandra} observations showed a wide temperature range indicative of shock heating of the circumstellar medium by the expanding shock wave \citep{DHN09}.

\citet{HKL07} used {\it Swift} observations of the super-soft source to derive a mass of the WD as 1.35 $\pm$ 0.01 M$_{\odot}$ and suggested that the mass of the WD is increasing at a rate of $\sim$~(0.5$-$1) $\times$ 10$^{-7}$ M$_{\odot}$ yr$^{-1}$.

\section{Final Thoughts}
Research on RS Oph is still ongoing and many questions remain unanswered. These include, for example:
\begin{itemize}
\item What happens to the ejecta as it progresses from the first {\it HST} epoch to the second? Is there a deceleration of the inner component while the outer moves freely? Or is there something else going on? What will this tell us about remnant shaping? This is possible to answer if spectra were available around the same time as the second {\it HST} epoch and applying the model from the first epoch. We can accept that the outer component arises from the expanding ejecta, but what about the inner one? Is it some post-outburst wind?
\item What is the nature of the supposed pre-outburst signal? This of course poses serious issues as in the TNR model the outburst is effectively instantaneous. So is there another mechanism at play before the TNR outburst for example, some increase in mass accretion rate before the outburst \citep[e.g.,][]{KP09}? Perhaps this could be resolved if RS Oph is observed to undergo dwarf novae outbursts during quiescence. 
\item What is the underlying composition of the WD? This is fundamental for our understanding of any hints to the progenitors of Type Ia SN. The short recurrence timescale and the duration of the super-soft source suggests a high mass WD. From stellar evolution, generally, a $>$ 1.1 M$_{\odot}$ WD is ONe, as opposed to CO, which will not explode as a Type Ia SN. U Sco, another RN with a similarly high mass WD, was strongly suggested to comprise an ONe WD \citep{M11}. This point requires special attention in order to fully understand the role of RNe as progenitors of Type Ia SNe. 
\end{itemize}

\begin{acknowledgements}
The author would like to thank Rebekah Hounsell and Stewart Eyres for providing the data for the SMEI light curve and modified version of Figure~\ref{fig:wave}, and Michael Bode for comments on a first draft. The author would also like to thank financial support from the Royal Astronomical Society and the SA SKA Office. 
\end{acknowledgements}

\balance

\bibliographystyle{aa}

\end{document}